\documentclass[nofootinbib,superscriptaddress,a4paper,twocolumn,longbibliography,dvipsnames]{revtex4-2}
\usepackage{graphicx}% Include figure files
\usepackage{dcolumn}% Align table columns on decimal point
\usepackage{bm}% bold math
\usepackage{subfigure}
\usepackage{physics}
\usepackage{tikz}
\usepackage{qcircuit}
\usepackage{booktabs}
\usepackage{algpseudocode}
\usepackage{colortbl}
\usepackage{soul} 
\usepackage{threeparttable}
\usepackage{mathrsfs}
\newsavebox{\imagebox}

\usepackage{hyperref}% add hypertext capabilities, should be last include
\usepackage{tikz}
\usepackage{calc}
\usepackage{newfloat}
\DeclareFloatingEnvironment[
    fileext=loa,
    listname={List of Algorithms},
    name=Algorithm,
    placement=tbhp,
    within=section,
]{algorithm}
\newcommand\mdoubleplus{\mathbin{+\mkern-10mu+}}

\usepackage{url}
\usepackage{xspace}
\usepackage{wrapfig}
\usepackage{booktabs}
\usepackage{rotating}

\usepackage{amssymb}% http://ctan.org/pkg/amssymb
\usepackage{pifont}% http://ctan.org/pkg/pifont
\graphicspath{
	{pics/}
}
\DeclareUnicodeCharacter{200B}{}
\DeclareUnicodeCharacter{03B1}{}
\DeclareUnicodeCharacter{03B2}{}
\DeclareUnicodeCharacter{0308}{}
\begin{document}
% \begin{rotate}{90}
% \begin{sideways}
% \title{Quantum-classical algorithm generates lead inhibitor of KRAS G12D}
%\title{Quantum-classical algorithm generates  inhibitor of KRAS G12D}
%\title{Quantum-Classical Algorithm Unveils Novel Scaffolds Targeting KRAS Inhibition}
%\title{Quantum-Classical Algorithm Unveils Novel Inhibitors for KRAS}
\title{Quantum Computing-Enhanced Algorithm Unveils Novel Inhibitors for KRAS}

\author{Mohammad Ghazi Vakili}
\affiliation{Department of Computer Science, University of Toronto, Canada} 
\affiliation{Department of Chemistry, University of Toronto, Canada}

\author{Christoph Gorgulla}
\affiliation{Department of Structural Biology, St. Jude Children's Research Hospital, USA} 
\affiliation{Department of Physics, Harvard University, USA}
\author{AkshatKumar Nigam}
\affiliation{Department of Computer Science, Stanford University} 

\author{Dmitry Bezrukov}
\affiliation{Insilico Medicine AI Limited, Abu Dhabi, UAE}

\author{Daniel Varoli}
\affiliation{Zapata AI, 100 Federal St., Boston, MA 02110} 

\author{Alex Aliper}
\affiliation{Insilico Medicine AI Limited, Abu Dhabi, UAE}
\author{Daniil Polykovsky}
\affiliation{Insilico Medicine AI Limited, Abu Dhabi, UAE} 

\author{Krishna M. Padmanabha Das}
\affiliation{Department of Cancer Biology, Dana-Farber Cancer Institute, Boston, MA, USA}
\affiliation{Department of Biological Chemistry and Molecular Pharmacology, Harvard Medical School, Harvard University, Boston, MA, USA}

\author{Jamie Snider}
\affiliation{Donnelly Centre, Temerity, Faculty of Medicine, University of Toronto, ON, Canada} 

\author{Anna Lyakisheva}
\affiliation{Donnelly Centre, Temerity, Faculty of Medicine, University of Toronto, ON, Canada} 

\author{Ardalan Hosseini Mansob}
\affiliation{Donnelly Centre, Temerity, Faculty of Medicine, University of Toronto, ON, Canada}
\affiliation{Department of Molecular Genetics, University of Toronto, Ontario, Canada}

\author{Zhong Yao}
\affiliation{Donnelly Centre, Temerity, Faculty of Medicine, University of Toronto, ON, Canada} 
\author{Lela Bitar}
\affiliation{Donnelly Centre, Temerity, Faculty of Medicine, University of Toronto, ON, Canada}
\affiliation{Department for Lung Diseases Jordanovac, Clinical Hospital Centre Zagreb, University of Zagreb, Croatia} 

% \author[3]{\fnm{Eugene} \sur{Radchenko}}
\author{Eugene Radchenko}
\affiliation{Insilico Medicine AI Limited, Abu Dhabi, UAE} 

\author{Xiao Ding}
\affiliation{Insilico Medicine AI Limited, Abu Dhabi, UAE} 

% \author[3]{\fnm{Jinxin} \sur{Liu}}
\author{Jinxin Liu}
\affiliation{Insilico Medicine AI Limited, Abu Dhabi, UAE} 

% \author[3]{\fnm{Fanye} \sur{Meng}}
\author{Fanye Meng}
\affiliation{Insilico Medicine AI Limited, Abu Dhabi, UAE} 

% \author[3]{\fnm{Feng} \sur{Ren}}
\author{Feng Ren}
\affiliation{Insilico Medicine AI Limited, Abu Dhabi, UAE} 

% \author[4]{\fnm{Yudong} \sur{Cao}}
\author{Yudong Cao}
\affiliation{Zapata AI, 100 Federal St., Boston, MA, USA} 

% Igor Stagljar
\author{Igor Stagljar}
\affiliation{Donnelly Centre, Temerity, Faculty of Medicine, University of Toronto, ON, Canada}
\affiliation{Department of Biochemistry, University of Toronto, Ontario, Canada}
\affiliation{Department of Molecular Genetics, University of Toronto, Ontario, Canada}
\affiliation{Mediterranean Institute for Life Sciences, Split, Croatia}

\author{Al\'an Aspuru-Guzik}
\email{Correspondence to: aspuru@utoronto.ca}
\affiliation{Department of Computer Science, University of Toronto, Canada}
\affiliation{Department of Chemistry, University of Toronto, Canada}
\affiliation{Department of Chemical Engineering and Applied Chemistry, University of Toronto, Canada}
\affiliation{Department of Materials Science and Engineering, University of Toronto, Canada}
\affiliation{Vector Institute for Artificial Intelligence, Toronto, Canada}
\affiliation{Lebovic Fellow, Canadian Institute for Advanced Research (CIFAR), Toronto, Ontario, Canada}

% \author[3]{\fnm{Alex} \sur{Zhavoronkov}}
\author{Alex Zhavoronkov}
\email{Correspondence to: alex@insilicomedicine.com}
\affiliation{Insilico Medicine AI Limited, Abu Dhabi, UAE}

\begin{abstract}
\textbf{Abstract}
The discovery of small molecules with therapeutic potential is a long-standing challenge in chemistry and biology. Researchers have increasingly leveraged novel computational techniques to streamline the drug development process to increase hit rates and reduce the costs associated with bringing a drug to market. To this end, we introduce a quantum-classical generative model that seamlessly integrates the computational power of quantum algorithms trained on a 16-qubit IBM quantum computer with the established reliability of classical methods for designing small molecules. Our hybrid generative model was applied to designing new KRAS inhibitors, a crucial target in cancer therapy. We synthesized 15 promising molecules during our investigation and subjected them to experimental testing to assess their ability to engage with the target. Notably, among these candidates, two molecules, ISM061-018-2 and ISM061-22, each featuring unique scaffolds, stood out by demonstrating effective engagement with KRAS. ISM061-018-2 was identified as a broad-spectrum KRAS inhibitor, exhibiting a binding affinity to KRAS-G12D at $1.4 \mu M$. Concurrently, ISM061-22 exhibited specific mutant selectivity, displaying heightened activity against KRAS G12R and Q61H mutants. To our knowledge, this work shows for the first time the use of a quantum-generative model to yield experimentally confirmed biological hits, showcasing the practical potential of quantum-assisted drug discovery to produce viable therapeutics. Importantly, comparative analysis with existing classical generative models indicates that integrating quantum computing enhances distribution learning from established datasets, suggesting a potential advantage for quantum generative models over their classical counterparts. Moreover, our findings reveal that the efficacy of distribution learning correlates with the number of qubits utilized, underlining the scalability potential of quantum computing resources. Overall, we anticipate our results to be a stepping stone towards developing more advanced quantum generative models in drug discovery.
\end{abstract}

\maketitle 
%%%%%%%%%%%%%%%%%%%%%%%%%%%%%%%%%MAIN%%%%%%%%%%%%%%%%%%%%%%%%%%%%%%%%%%%%%%%%%%%%

\begin{figure*}
  \centering
 \includegraphics[scale=1.4]{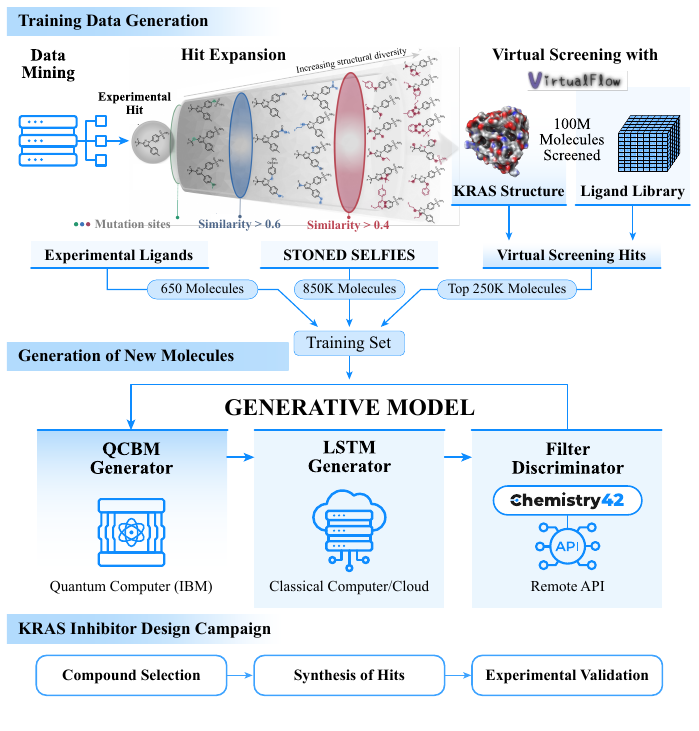}
\caption{\textbf{Schematic Representation of the Hybrid Quantum-Classical Framework for KRAS Ligand Development.} The initial phase concentrates on compiling a dataset for model training. A curated set of 650 experimentally verified inhibitors targeting the KRAS protein is extracted from the literature. By applying the STONED-SELFIES algorithm, analogs for each identified compound are derived, yielding an expanded collection of around 850,000 compounds. This dataset is further enhanced by the addition of the top 250,000 candidates, identified via a virtual screening process using the REAL ligand library against the KRAS protein, culminating in a dataset of over 1 million molecules for training our generative model. Upon completing the training of our model, new molecules targeting KRAS are created employing both a classical LSTM network and a Quantum Circuit Born Machine (QCBM) as the underlying generative frameworks. The LSTM network processes sequential data encapsulating the chemical structures of ligands, while QCBM, trained based on the quality of LSTM-generated samples, creates complex, high-dimensional probability distributions. The combined workflow utilizes Chemistry42 as a reward function to incentivize the creation of structurally diverse and synthesizable molecules. 
}
  \label{fig:workflow}
\end{figure*}

\section{Introduction}\label{chp:intro}
Drug discovery is a multifaceted process involving various stages, including identifying, developing, and rigorously testing novel molecules intended to combat a spectrum of diseases \cite{petrova2013innovation}. A drug discovery campaign typically spans a decade to fifteen years or more and commands a  financial commitment that often exceeds \$2.5 billion \cite{dimasi2016innovation} during clinical trials. Significantly, these substantial investments do not guarantee success; when a drug development cycle fails, it represents a  financial setback with the potential loss of the entire capital investment \cite{sun202290}. Consequently, the pharmaceutical industry continually seeks innovative and cutting-edge technologies to integrate into their workflows, aiming to enhance their prospects for successful market entry. \\

The drug discovery journey commences with identifying a critical target, usually a protein or enzyme integral to a disease's pathophysiology \cite{hughes2011principles}. Following this  step, researchers employ many techniques, notably virtual screening \cite{gorgulla2020open,gorgulla2023virtualflow, nigam2023drug}, to design and rigorously assess potential drug candidates creatively. These candidates are meticulously evaluated for their proficiency in engaging with and modulating the target, propelling the pursuit of therapeutic innovations \cite{paul2010improve}. Concurrently, generative modelling is emerging as a transformative technology in molecule design \cite{zhavoronkov2019deep,stokes2020deep,ren2023alphafold,nigam2024artificial}. Generative models utilize machine learning techniques to comprehend the underlying distribution of atoms and bonds in a specified dataset. Subsequently, these models are employed to construct molecules with predefined properties, a process known as inverse molecular design \cite{sanchez2018inverse, pollice2021data, nigam2021assigning}. A promising aspect of these models is their ability to navigate the vast chemical space, proposing interesting molecules within the challenging realm of $10^{60}$ drug-like molecules \cite{bohacek1996art}. \\

A transition from the intricate landscape of traditional drug discovery to the realm of advanced computational techniques underscores the industry's adaptation to innovative methodologies. Within this evolution, quantum machine learning has garnered attention, particularly in enhancing generative models. Hibat-Allah et al. \cite{hibat-allah_framework_2023} introduced a framework that juxtaposes the performance of quantum and classical generative models, focusing on the practical quantum advantage and the potential superiority of Quantum Circuit Born Machines (QCBM) over their conventional counterparts. Their research highlights the generalization capabilities of QCBMs, notably in generating novel, valid samples that extend beyond the training dataset. These models adhere to the target distribution and address training challenges such as barren plateaus, effectively integrating tensor networks alongside QCBMs \cite{mcclean2018barren, gili_quantum_2023}. Despite the promising strides, quantum information processing confronts inherent limitations, notably in data loading and trainability within quantum circuits. These challenges have prompted a shift towards hybrid algorithms, amalgamating the strengths of both quantum and classical machine learning paradigms. A notable contribution in this regard is from Manuel et al. \cite{rudolph_generation_2022}, who demonstrated the enhanced exploration of target space facilitated by integrating a multi-basis QCBM into a classical framework, termed the Quantum Circuit Associative Adversarial Network (QC-AAN). This hybrid model not only augments model training beyond the capabilities of classical hardware but also transcends the constraints imposed by the size of the quantum prior, signifying a significant leap in the field of quantum-enhanced machine learning. Furthermore, Zeng et al. \cite{zeng_conditional_2023} introduced a quantum-classical hybrid framework (conditional QCBM and a classical model) for image generation tasks, utilizing quantum circuits that encode conditional information via additional qubits. \\

Here, we propose a novel quantum-classical generative model specifically designed to overcome qubit limitations while combining the best of classical and quantum computational methods. Our hybrid model is engineered to generate realistic ligands to design compounds for targeted proteins. A prime focus of our endeavour is the KRAS protein, a target notorious for its intricate complexity and historical resistance in drug discovery ventures \cite{bank_rcsb_nodate, wang_identification_2022, mao_krasg12d_2022}. The comprehensive workflow of our approach, spanning from data generation to experimental validation, is illustrated in Figure \ref{fig:workflow}. Notably, our model capitalizes on established classical computational pipelines. We employ the STONED-SELFIES algorithm \cite{nigam_beyond_2021} for its proficiency in exploring and interpolating the vast chemical space. Additionally, we incorporate Chemistry42 \cite{ivanenkov_chemistry42_2023} as an external reward signal, a strategic move to enhance the realistic design of molecules. This integration ensures that our generated ligands exhibit novel and drug-like properties and are highly relevant for the targeted protein. To validate the predictions of our model, we synthesized a series of 15 candidate compounds. These engineered compounds underwent extensive experimental evaluation to ascertain their efficacy in targeting KRAS. One candidate, distinguished by a novel chemotype, manifested as a pan-KRAS inhibitor. We believe that this study lays foundational groundwork for the integration of quantum algorithms in the field of drug discovery. This work underscores the potential and viability of quantum-enhanced methodologies in drug discovery by achieving what we understand to be the first experimental hit attributed to a quantum algorithm.

\begin{figure*}
    \centering
    \includegraphics[width=\linewidth]{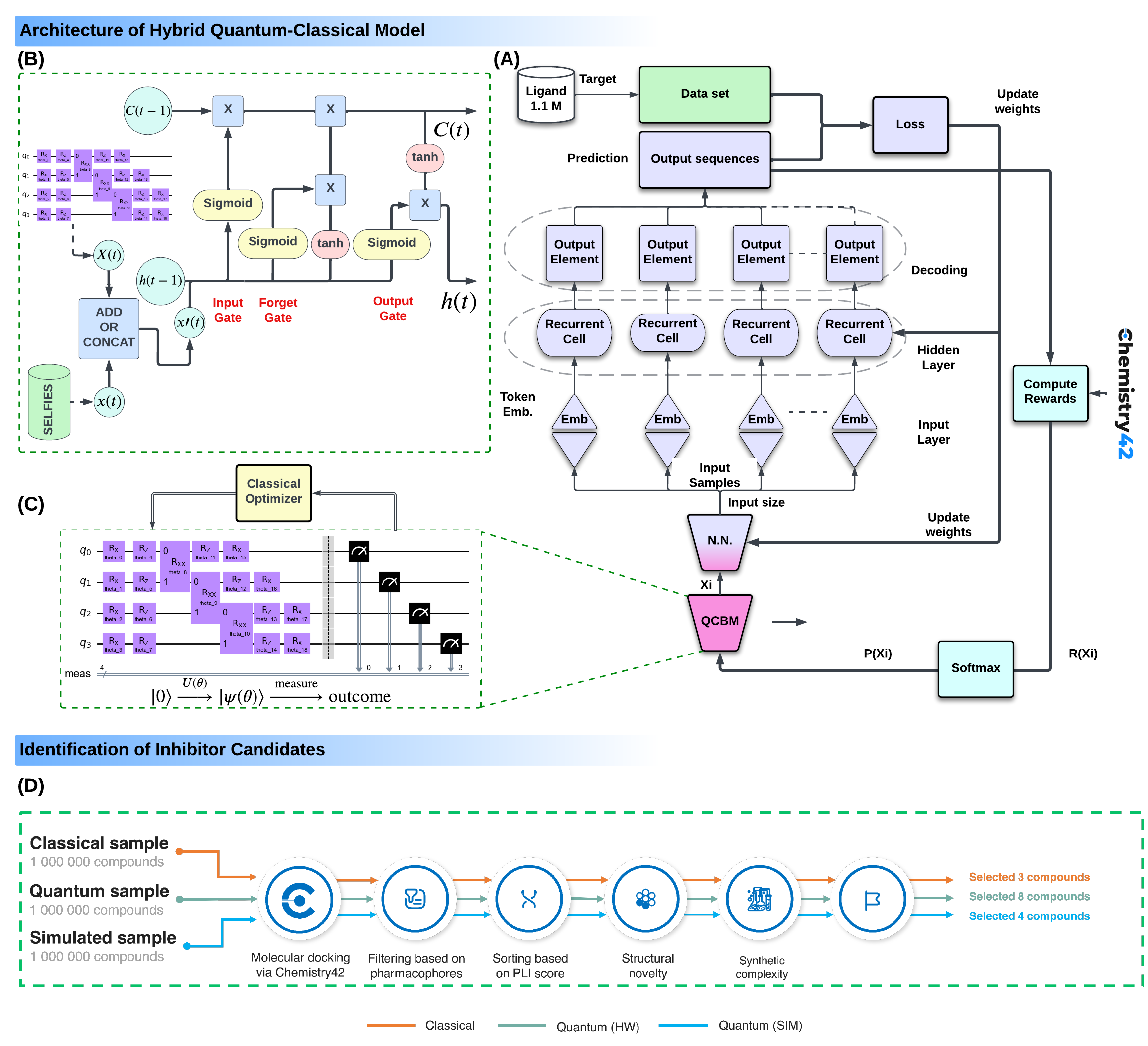}
    % {figures/fig2_reworked.png}
    \caption{\textbf{Quantum-Enhanced Generative Model for Drug Discovery Applications.} \textbf{(A)} Hybrid model combining a Quantum Circuit Born Machine (QCBM) with Long Short-Term Memory (LSTM). This model iteratively trains using prior samples from quantum hardware. \textbf{(B)} Integration method of prior samples into the LSTM architecture. Molecular information (in SELFIES encoding) and quantum data are merged by addition or concatenation. The resultant samples, \(X'(t)\), are then input to the LSTM cell. \textbf{(C)} Quantum prior component described as a QCBM, generating samples from quantum hardware each training epoch and trains with a reward value, \(P(x) = \text{Softmax}(R(x))\), calculated using Chemistry42 or a local filter. \textbf{(D)} Process of experimental sample selection: 1 million compounds are sampled from each model—classical samples (via vanilla LSTM), quantum samples (QCBM on quantum hardware), and simulated samples (quantum simulation on classical hardware). These samples undergo evaluation by Chemistry42, filtering out compounds unsuitable for pharmacological purposes and ranking the remaining compounds by their docking score (PLI score). Subsequently, 15 novel compounds were selected for synthesis.}
    \label{fig:qegm}
\end{figure*}

\begin{figure*}
    \centering
    \includegraphics[width=\linewidth]{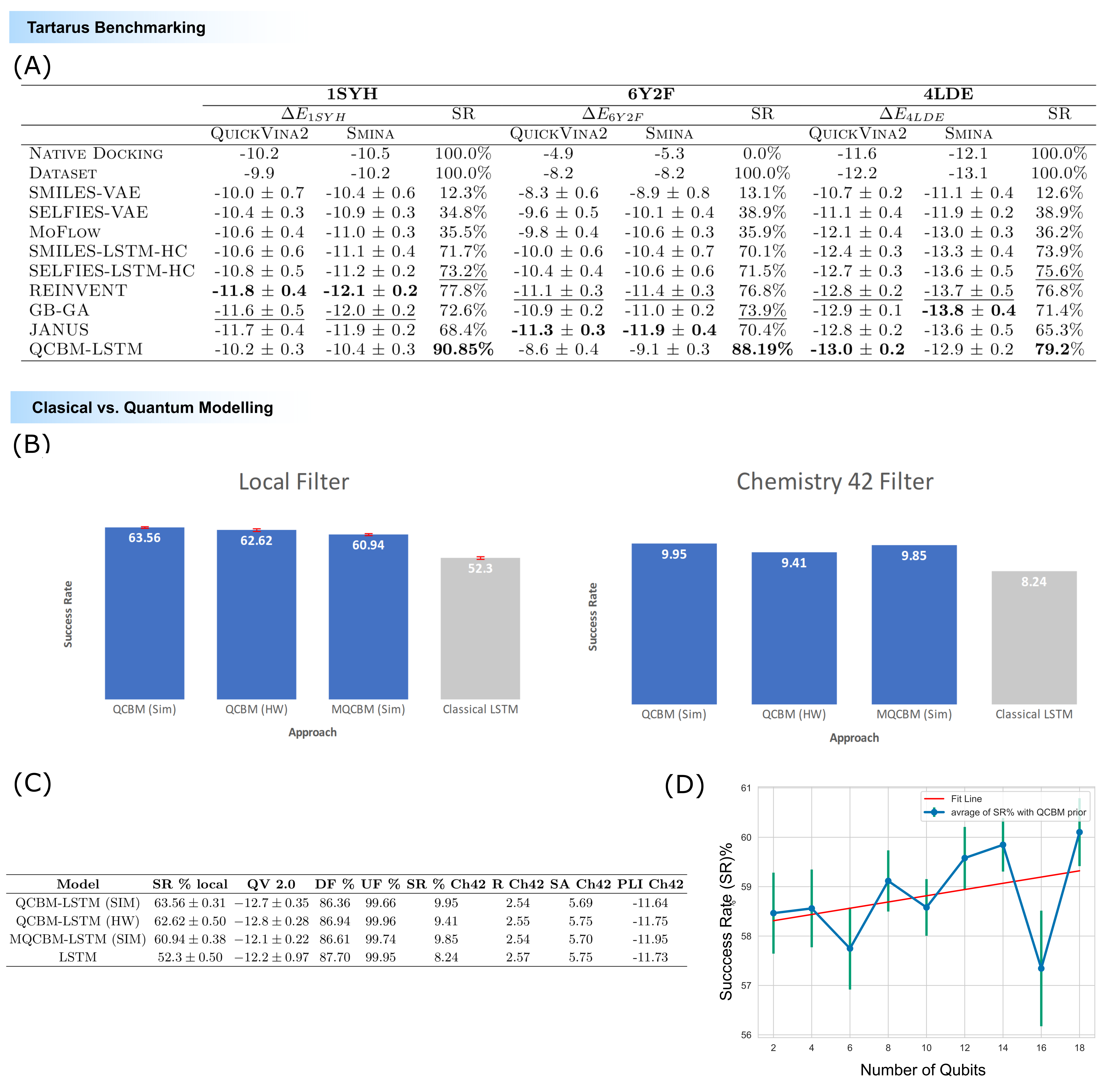}
\caption{\textbf{Comparative Benchmarking of Quantum and Classical Ligand Design Methods.} \textbf{(A)} Evaluation of the proposed model against classical counterparts using the Tartarus benchmark suite \cite{nigam_tartarus_2023} for ligand design across three protein targets: 1SYH, 6Y2F, 4LDE, with models trained on a subset of the DTP Open Compound Collection. Displayed metrics show both the average and the variability (mean ± standard deviation) of the optimal objective values for the targets, compiled from five individual experiments. 'Dataset' refers to the molecule with the highest performance in the training dataset, whereas 'Native Docking' indicates the initial ligands within their crystallographic structures. The notation $\Delta E_{X}$ signifies the docking score relative to the protein target designated by $X$. SR stands for the success ratio, indicating the percentage of molecules that meet the predefined structural benchmarks. \textbf{(B)} Comparative analysis of our hybrid approaches with varied priors. The performance of the Quantum Circuit Born Machine (QCBM) was assessed using both a quantum simulator (Sim) and a hardware backend (HW), and contrasted with a Multi-bases QCBM (MQCBM) operating solely on a quantum simulator (SIM), as well as an LSTM model devoid of quantum priors (representing a fully classical architecture). We calculated the number of generated molecules that met a series of synthesizability and stability criteria as stipulated by the Tartarus benchmarking platform (referred to as Local Filters) and by Chemistry42 (referred to as Chemistry42 Filters).  \textbf{(C)} Comparative analysis of prior sampling techniques in producing high-docking molecules, as assessed by QuickVina2 and Chemistry42. This comparison delineates the Success Rate (SR \%) of molecules meeting Tartarus filter criteria, the uniqueness of generated ligands (Unique Fraction, UF \%), and the Structural Diversity Fraction (DF \%) of the generated ligands, across various methods.  the success rates (SR\% Ch42) of molecules meeting Chemistry42's filter criteria, the top reward values (R Ch42) assigned to molecules by Chemistry42, the synthetic accessibility score (SA Ch42) of drug-like molecules, and the highest PLI Ch42 scores found in the generation. The PLI score is measured in kcal/mol, with more negative values indicating better scores. \textbf{(D)} Success rate of generating molecules that meet Tartarus's filter criteria as a function of the number of qubits used in modeling priors for the QCBM.}
    \label{fig:comparison}
\end{figure*}

\section{Results and Discussion}\label{chp:results}

 Our methodology encompasses a comprehensive workflow, extending from data preparation to experimental validation, as delineated in Figure \ref{fig:workflow}. This workflow is structured into three pivotal stages:\\

\textbf{\textit{(1)} Generation of Training Data}: We initiate the process by constructing a robust dataset for training our generative model to target the KRAS protein. The foundation of this dataset is approximately 650 experimentally confirmed KRAS inhibitors, compiled through an extensive literature review \cite{kwan2022path, parikh2022drugging, srisongkram2023prediction, nagasaka2020kras}. Acknowledging the necessity of a more expansive dataset to develop a model for ligand design effectively, we adopted a two-pronged approach: virtual screening and local chemical space exploration. In the virtual screening phase, we employed Virtual Flow 2.0 \cite{gorgulla2023virtualflow} to screen 100 million molecules, utilizing Enamine's REAL library \cite{grygorenko2020generating} in conjunction with molecular docking techniques. The top 250k compounds from this screen, exhibiting the lowest docking scores, were subsequently integrated into our dataset. Complementing this, the local chemical space exploration was conducted using the STONED-SELFIES algorithm \cite{nigam2021beyond}, which was applied to the 650 experimentally derived hits. This algorithm distinctively introduces random point mutations into the SELFIES representations \cite{krenn2020self, lo2023recent, krenn2022selfies} of the molecules, thereby generating novel compounds that maintain a structural resemblance to the starting point. The resulting derivatives were filtered based on synthesizability (see Methods \ref{appx:stonedselfies} and \ref{appx:virtualflow} for details), culminating in the addition of 850 thousand molecules to our training set. \\

\textbf{\textit{(2)} Generation of New Molecules}: Our approach is structured around the integration of three primary components: a) the Quantum Circuit Born Machine (QCBM), b) the classical Long Short Term Memory (LSTM) model, and c) Chemistry42 for AI-driven validation, as shown in Figure \ref{fig:qegm}. The QCBM generator \cite{hibat-allah_framework_2023} employs a 16-qubit IBM quantum processor, utilizing quantum circuits to model complex data distributions. The integration method of quantum priors into the LSTM architecture, as shown in Figure \ref{fig:qegm}B, involves merging molecular information encoded in SELFIES and quantum data by addition or concatenation to form samples, \(X'(t)\), which are then input into the LSTM cell. The quantum component, depicted in Figure \ref{fig:qegm}C, is a QCBM that generates samples from quantum hardware each training epoch and is trained with a reward value, \(P(x) = \text{Softmax}(R(x))\), calculated using Chemistry42 or a local filter. This cyclical sampling, training, and validation process forms a loop aimed at continually improving the generated molecular structures for targeting KRAS. \\

\textbf{\textit{(3)} Experimental validation}: The process of selecting experimental sample candidates is illustrated in Figure \ref{fig:qegm}D. After training our model, we sampled 1 million compounds from each prior model listed in Figure \ref{fig:qegm}D. These samples underwent evaluation by Chemistry42, filtering out unsuitable compounds for pharmacological purposes and ranking the remaining compounds by their docking score (PLI score). Subsequently, 15 novel compounds were selected for synthesis and underwent Surface Plasmon Resonance (SPR) and cell-based assay experiments. \\

%The results obtained from the classical and quantum simulations conducted in our study are presented and discussed in this section. The quantum-classical approach demonstrated a high success rate in computational benchmark studies targeting KRAS G12D, outperforming a similar classical model. Moreover, a benchmark comparison with state-of-the-art models verified the capability of the proposed model. Selected compounds generated by this novel approach have undergone experimental validation and synthesis, indicating the practical applicability of the hybrid generative model in drug discovery. \\

Before initiating our campaign to design new KRAS inhibitors, we sought to compare our hybrid quantum-classical approach with established classical algorithms. Two principal questions guided our evaluation: First, does integrating quantum methodologies contribute to generating high-quality molecules featuring potentially strong target-specific docking scores? To address this, we evaluated our approach against the Tartarus benchmarking suite \cite{nigam_tartarus_2023}, specifically designed for drug discovery tasks on three distinct protein targets. Second, we examined the influence of the quantum prior, replacing it with a classical counterpart and increasing the number of qubits in our quantum model to determine whether there is a proportional relationship between the number of qubits and the quality of the generated molecules. 

\subsection{Computational Benchmarks - Classical vs Quantum Models}

\subsubsection{Tartarus Benchmark}
We employed the Tartarus platform \cite{nigam_tartarus_2023} to benchmark our proposed QCBM-LSTM methodology against an array of classical state-of-the-art models, including REINVENT \cite{olivecrona2017molecular}, SMILES-VAE \cite{G_mez_Bombarelli_2018}, SELFIES-VAE \cite{krenn2022selfies}, MoFlow \cite{zang2020moflow}, SMILES-LSTM-HC \cite{segler2018generating,brown2019guacamol}, SELFIES-LSTM-HC, GB-GA \cite{jensen2019graph}, and JANUS \cite{nigam2022parallel}. The study focused on three protein targets selected from the Tartarus dataset: 1SYH, an ionotropic glutamate receptor associated with neurological and psychiatric disorders such as Alzheimer's, Parkinson's, and epilepsy \cite{frandsen_tyr702_2005}; 6Y2F, the main protease of the SARS-CoV-2 virus, crucial for its RNA translation \cite{zhang_crystal_2020}; and 4LDE, the $\beta_2$-adrenoceptor GPCR, a cell membrane-spanning receptor that binds to adrenaline, a hormone implicated in muscle relaxation and bronchodilation \cite{ring_adrenaline-activated_2013}. For each target, we had a dual objective: to generate novel molecules that exhibit strong binding affinity to the specified proteins, as determined by active sites assigned by Tartarus, and to minimize the docking score using QuickVina2 \cite{alhossary2015fast}. Additionally, these molecules were required to pass a comprehensive set of filters designed to eliminate reactive, unsynthesizable, or unstable groups, thereby streamlining the drug discovery process. The top-performing molecules, post-filtering, were subject to a refined re-scoring using a more precise scoring function provided by SMINA \cite{koes2013lessons}, at an increased level of exhaustiveness.\\

We conducted experiments utilizing the QCBM with 16 qubits as a quantum prior and the LSTM as a classical model. The local filter from the Tartarus paper served as the reward function to train the QCBM. As recommended by Tartarus, our models were trained on a subset of 150,000 molecules from the Developmental Therapeutics Program (DTP) Open Compound Collection \cite{voigt2001comparison, ihlenfeldt2002enhanced}, referred to as \texttt{DATASET} in Figure \ref{fig:comparison}, Table (A). Notably, all 150,000 structures underwent a rigorous screening process using structural filters to eliminate reactive, unsynthesizable, or unstable groups. As such, generative models adept at capturing the distribution of the provided molecule set would exhibit a correspondingly high success rate in generating novel molecules without structural violations. Our observations indicate that only a few generative models demonstrate a high success rate. However, QCBM-LSTM model is very strong in producing a substantial number of high-quality samples that successfully meet the filter criteria, as evidenced by the elevated success rate (SR) depicted in Figure \ref{fig:comparison}, Table (A). Consequently, we believe that the incorporation of a quantum prior leads to improved distribution matching. We further benchmark the influence of a classical/quantum prior in the subsequent section. Moreover, our analysis reveals that for the 4LDE target, our model generates the highest-scoring molecules relative to other generative models. While the docking scores for the remaining two targets are not as high as those produced by classical algorithms, we speculate that incorporating a docking-score-based reward, in conjunction with the filter success rate, could potentially improve our results. 

\subsubsection{Benchmarking of Prior distributions}
To evaluate the impact of prior selection on the quality of the molecules generated by our model, we trained four distinct model variants, each incorporating different priors (refer to Figure \ref{fig:comparison}(B)). Specifically, we examined a Quantum Circuit Born Machine (QCBM) prior and implemented it on both a quantum simulator (Sim) and a hardware backend (HW), in contrast with a Multi-bases QCBM (MQCBM) operating exclusively on a quantum simulator (Sim), and a classical LSTM model devoid of quantum priors. These models were tasked with designing KRAS inhibitors, utilizing a meticulously curated dataset of over one million molecules (see Figure \ref{fig:workflow}). Figure \ref{fig:comparison}(B) showcases the optimal results obtained following a comprehensive optimization of the corresponding architectures using Optuna \cite{akiba2019optuna} (detailed in Methods \ref{appx:benchmark}). We assessed the quality of the generated molecules employing two distinct sets of criteria: one derived from Tartarus \cite{nigam_tartarus_2023}, termed the "Local Filter," and a more stringent set provided by Chemistry42, termed the "Chemistry42 Filter." In both assessments, we observed that incorporating a quantum prior  enhances the success rate, as gauged by the proportion of molecules satisfying the criteria set by the two filters. Furthermore, utilizing the top model from each prior category, we sampled 5,000 molecules that successfully met the filter criteria and examined their respective docking scores (as presented in Figure \ref{fig:comparison}, Table (C)). Intriguingly, these molecules displayed comparably high docking scores as determined by QuickVina2 (denoted as QV 2.0 in the Table) and the protein-ligand interaction score (PLI), as evaluated by Chemistry42 (noted as PLI Ch42 in the Table). Additionally, the synthesized molecules demonstrated consistent metrics across various parameters, including Diversity Fraction (DF\%), Uniqueness Fraction (UF\%), Chemistry42 Reward (CH42 R), and the Chemistry42 Synthetic Accessibility Score (Chemistry42 SA) \cite{ivanenkov_chemistry42_2023}.\\
% (detailed in Methods \ref{TODO}).\\

Encouraged by our observation that quantum priors enhance molecule quality, we further investigated the influence of the number of qubits used in modelling priors on the quality of generated molecules, as shown in Figure \ref{fig:comparison}(D). Specifically, we analyzed the percentage of 5,000 uniquely generated random molecules that satisfied a series of local filters. Interestingly, our findings reveal that the success rate correlates roughly linearly with the number of qubits employed in modelling the prior, indicating a direct relationship between the complexity of the quantum model and the effectiveness in generating high-quality molecules. This trend underscores the potential of increasing qubit numbers in quantum models to improve molecular design outcomes systematically.

\begin{figure*}
    \centering
    \includegraphics[width=0.85\linewidth]{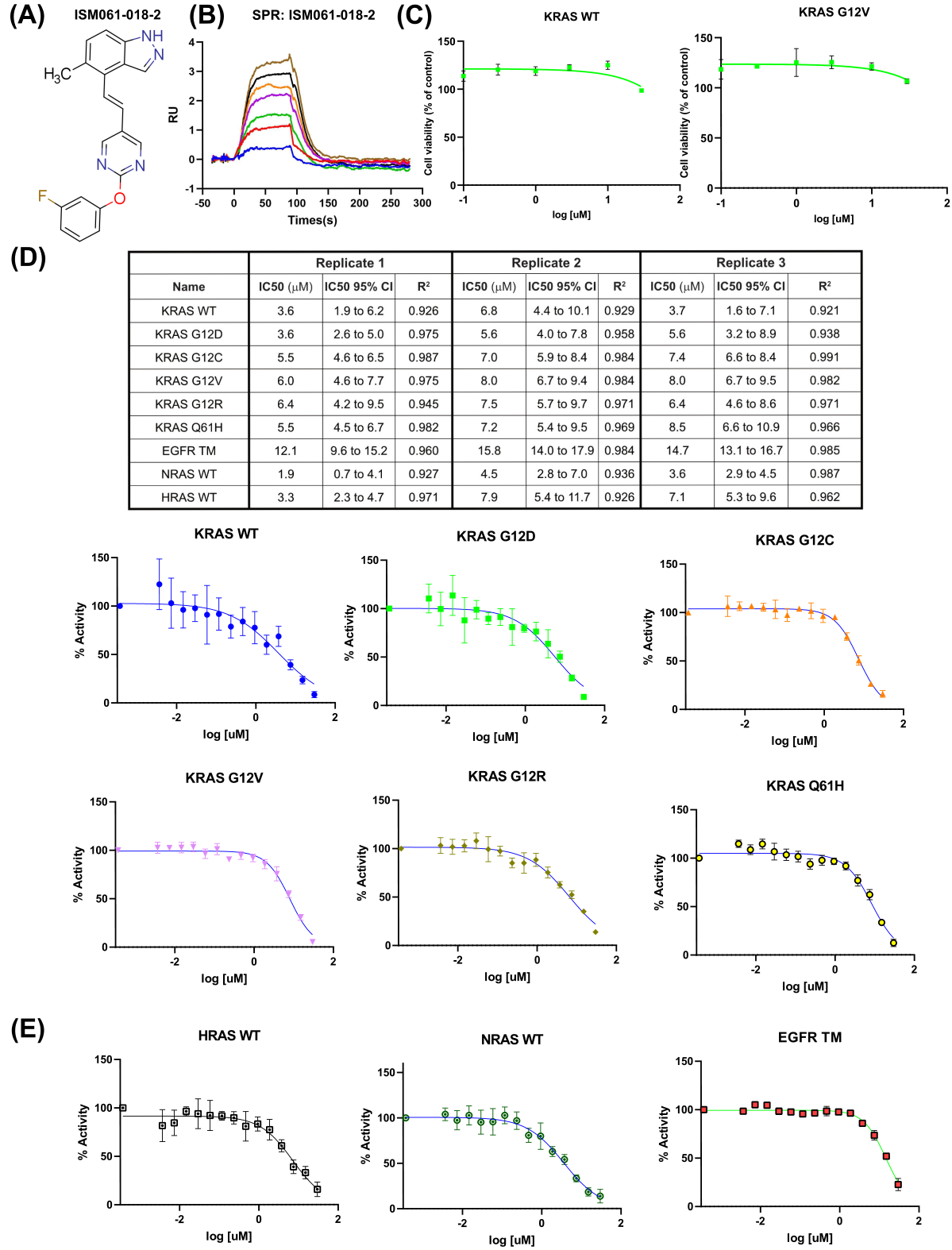}
\caption{\textbf{Pharmacological Characterization of Compound ISM061-018-2 Through Surface Plasmon Resonance and Cellular Activity Assays.} \textbf{(A)} Chemical structure of ISM061-018-2. \textbf{(B)} Surface Plasmon Resonance (SPR) sensorgram illustrating the binding kinetics of ISM061-018-2 with various KRAS proteins. \textbf{(C)} Results of Cell-Titer-Glo viability assays illustrating the impact of the compound on cellular proliferation across a concentration range from 123 nM to 30 µM. Reported values represent the mean of three technical replicates, with standard deviation (S.D.) indicated. \textbf{(D, Table)} A compendium of IC50 values derived from MaMTH-DS dose-response assays, conducted in biological triplicate, evaluating a range of RAS protein baits interactions with the RAF1 prey partner. Investigated RAS members include the wild-type forms of KRAS, HRAS, and NRAS, alongside five oncogenic KRAS mutants of clinical significance. The interaction between EGFR and the SHCI adapter was additionally examined as an off-target control. We provide 95\% confidence intervals and R-squared values to verify the accuracy of the curve fitting. \textbf{(D,E)}: Dose-response curves from MaMTH-DS assays graphing the modulation of activity of various KRAS proteins, NRAS, HRAS, and EGFR, in response to increasing concentrations of ISM061-018-2 (from 4 nM to 30 µM), plotted on a logarithmic scale. The curves displayed represent one set from three biological replicates. Each point denotes the mean of three to four technical replicates, with S.D. provided. Curve fitting was executed in GraphPad Prism as delineated in the Methods section. These profiles underscore the compound's differential potency against distinct targets, shedding light on its pharmacological spectrum.}
    \label{fig:fig_4}
\end{figure*}

\subsection{KRAS Inhibitor Design Campaign}
A critical aspect of our work involved validating the most promising compounds identified by our hybrid quantum-classical model, specifically targeting KRAS proteins. This validation process is crucial for substantiating our computational predictions and bridging the gap between theoretical models and their practical, real-world applications. Through the testing of these selected compounds in a campaign aimed at KRAS proteins, we aim to showcase the potential practicality of our proposed generative model, thereby emphasizing the real-world relevance of our findings in the field of drug discovery.

\subsubsection{Chemistry42 Post-Screening and Selection of Promising Candidate Structures for Synthesis}

Our study employed the Chemistry42 platform's structure-based drug design (SBDD) screening workflow to evaluate the generated structures and select promising ligand structures for KRAS inhibition. This comprehensive workflow includes a series of filters and scoring modules designed to efficiently identify structures with favourable drug-likeness, synthetic accessibility, and target interactions, as detailed in several steps (see Figure~\ref{fig:qegm}D).

The initial phase of screening involved the application of various 2D and 3D filters:
\begin{itemize}
\item 2D structural filters evaluating simple structural and compositional parameters, including hydrogen bond donors, oxygen atoms, aromatic atom fraction, and rotatable bonds.
\item 2D property filters assessing estimated compound properties like molecular weight, lipophilicity, topological polar surface area, and molecular flexibility.
\item Medicinal chemistry filters identifying undesirable or problematic structural fragments.
\item A synthetic accessibility filter based on the Retrosynthesis-Related (ReRSA) model \cite{zagribelnyy_retrosynthesis-related_2021}.
\item 3D pharmacophore hypothesis derived from the X-ray structure of the complexed inhibitor molecule (PDB: 7EW9) \cite{bank_rcsb_nodate}, indicating common structural features of known KRAS inhibitors (Figure~\ref{fig:SAR}).
\item Pocket-ligand interaction (PLI) score, based on molecular docking, to assess binding features of KRAS inhibitors (Figure~\ref{fig:SAR}).
\item Calculation of an integrated reward value based on the weighted scores.
\end{itemize}

For the final selection of candidate structures for synthesis, we imposed more stringent criteria to increase the likelihood of identifying compounds with the desired biological activity. These criteria, more restrictive than those used during the screening evaluation, included:
\begin{itemize}
\item Successful passage through all Chemistry42 filters.
\item An integrated reward value greater than 0.7.
\item A Protein-Ligand Interaction (PLI) score less than -8 kcal/mol.
\item A pharmacophore match score exceeding 0.7.
\item A Synthetic Accessibility (ReRSA) score below 5.
\end{itemize}

At the stage of final selection and determination of structures for subsequent experimental verification, the resulting sets of molecules at the last step were first clustered by chemical similarity, and 15–20 clusters were determined for each generation. Then, within each cluster, a ranking was carried out based on the ReRSA estimate and the Chemistry42 Protein-Ligand Interaction (PLI) score. Medical chemistry experts and 10 structures analyzed the resulting sets of 100-150 molecules, which were selected based primarily on chemical novelty, structural complexity and potentially undesirable chemical functionality. 

\subsubsection{Experimental Evaluation of Generated Compounds}
\begin{figure*}
    \centering
    \includegraphics[width=0.80\linewidth]{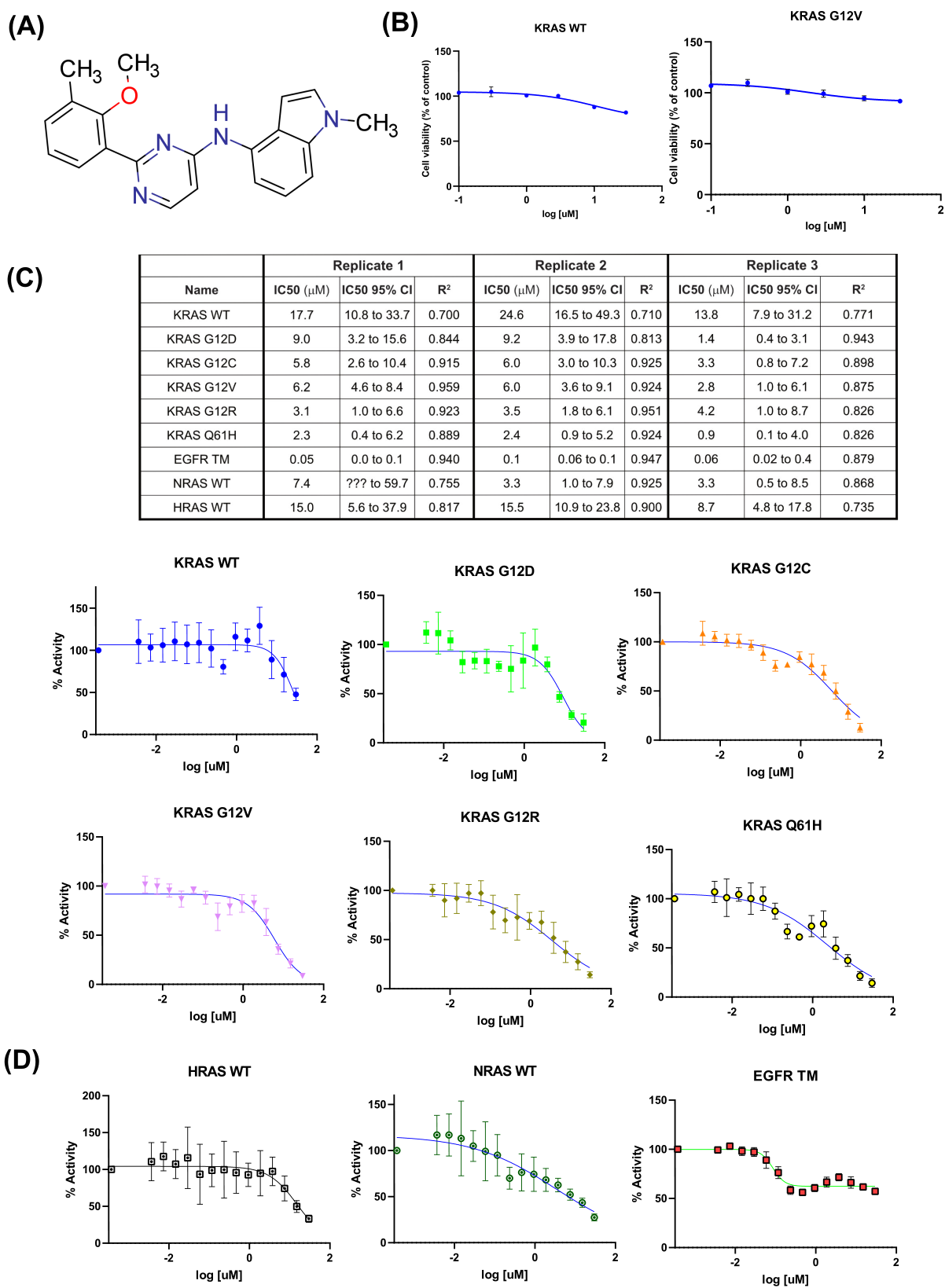}
\caption{\textbf{Pharmacological Evaluation of Compound ISM061-22 Against KRAS Variants and Other Related Proteins.} (A) Chemical structure of ISM061-22.
\textbf{(B)} Results from Cell-Titer-Glo viability assays depicting the effect of the compound on cellular proliferation over a concentration range from 123 nM to 30 µM. The values represent the mean of three technical replicates, with standard deviation (S.D.) indicated. 
\textbf{(C, Table)} Summary of IC50 values derived from MaMTH-DS dose-response assays conducted in biological triplicate, against various RAS protein baits' interactions with the RAF1 prey partner. Tested RAS members include wild-type KRAS, HRAS, and NRAS, as well as five clinically significant oncogenic KRAS mutants. The EGFR's interaction with the SHC1 adapter was also examined as an off-target control. The 95\% confidence intervals and R-squared values are reported to confirm the precision of the curve fitting. 
\textbf{(C,D)} Dose-response curves from MaMTH-DS assays, illustrating the modulation of activity of different KRAS proteins, NRAS, HRAS, and EGFR, by increasing concentrations of ISM061-22 (from 4 nM to 30 µM), presented on a logarithmic scale. The curves represent one instance from three biological replicates. Each data point is the average of three to four technical replicates, with S.D. presented. Curve fitting procedures were executed in GraphPad Prism, as outlined in the Methods section. The data emphasizes the compound's nuanced effectiveness against various protein targets, illuminating its potential therapeutic value.
}
\label{fig:fig_5}
\end{figure*}

From the pool of identified structures, we synthesized and characterized 15 compounds. Detailed methodologies of this process are elaborated in the Supplementary Materials\footnote{https://github.com/aspuru-guzik-group/quantum-generative-models} (Section S1). We showcased the molecular structures of the two most promising compounds (named ISM061-018-2 and ISM061-22) in Figures \ref{fig:fig_4}(A) and \ref{fig:fig_5}(A). Each synthesized compound underwent a rigorous two-phase evaluation: their binding affinities were determined using Surface Plasmon Resonance (SPR), and their biological efficacies were gauged through cell-based assays. Notably, the compound ISM061-018-2, engineered through our hybrid quantum model and illustrated in Figure \ref{fig:fig_4}, demonstrated a substantial binding affinity to KRAS G12D, registered at $1.4 \mu M$. To delve deeper into this molecule's effectiveness across a spectrum of KRAS mutations, we commenced an extensive series of tests using a cell-based assay. Specifically, we evaluated the molecule’s performance in a biological context, employing a commercial cell viability assay (CellTiter-Glo; Promega) in conjunction with MaMTH-DS, an advanced split-ubiquitin-based platform for the real-time detection of small molecules targeting specific cellular interactions \cite{saraon_drug_2020,stagljar1998genetic,petschnigg2014mammalian,saraon2021chemical,scheper2003coordination,benleulmi2016protein,lopes2015role}.\\

Further, the biological efficacy of ISM061-018-2 was rigorously tested. Interestingly, it demonstrated no detrimental impact on the viability of HEK293 cells, even when expressing either KRAS WT or KRAS G12V bait in MaMTH-DS format, and subjected to concentrations as high as $30 \mu M$ for 18-20 hours (Figure~\ref{fig:fig_4}C). Subsequent testing using MaMTH-DS across a spectrum of cell lines expressing various KRAS “baits” (both WT and five clinically significant oncogenic mutants) in combination with RAF1 “prey” (a recognized KRAS effector) revealed a dose-responsive inhibition of interactions, with $\mathrm{IC}{50}$s in the micromolar range (Figure~\ref{fig:fig_4}D). The compound's activity was not specific to mutants, as it targeted both WT and mutant interactions with similar efficacy. It also showed comparable effectiveness in disrupting the interactions of WT NRAS and HRAS “baits” with RAF1 “prey” (Figure~\ref{fig:fig_4}D,E). However, it was notably less potent, by a factor of 2-3 (based on $\mathrm{IC}{50}$ comparisons), against our control interaction comprising EGFR triple mutant “bait” and SHCI adapter “prey”, an interaction crucial in signalling but functioning upstream of RAS in the pathway (Figure~\ref{fig:fig_4}D,E). These results collectively hint at the potential pan-RAS activity of ISM061-018-2, albeit with indications of some off-target effects.\\

ISM061-22, as illustrated in Figure~\ref{fig:fig_5}A, also stood out as a compound of promise, particularly due to its selectivity towards certain KRAS mutants. This compound demonstrated a mild impact on cellular viability at high concentrations after 18-20 hours of exposure, as shown in Figure~\ref{fig:fig_5}C. Mirroring the performance of ISM061-018-2, ISM061-22 revealed dose-responsive inhibition of the interaction between KRAS "bait" and RAF1 "prey" within the micromolar range. Distinctively, it exhibited a markedly greater effect on mutant KRAS compared to WT, with the degree of influence ranging from 2-8 fold, depending on the specific mutant. Notably, KRAS G12R and Q61H displayed the highest sensitivity, as depicted in Figure~\ref{fig:fig_5}D,E. Diverging from ISM061-018-2, ISM061-22 did not show  binding to KRAS G12D. The compound also demonstrated activity against WT HRAS and NRAS, although it was less potent against HRAS. Furthermore, ISM061-22's interaction with the EGFR control displayed a unique inhibition profile, stabilizing at 50\% maximal activity instead of decreasing to zero, as typically observed in the RAS/RAF interaction curves (Figure \ref{fig:fig_5}D,E). This distinct pattern suggests an alternative mode of action for ISM061-22, potentially indicative of partial non-specific activity or a feedback mechanism influencing the EGFR signaling pathway.\\

In essence, these live-cell experimental observations underscore the robustness of our approach, effectively identifying small molecule candidates with biological activity. This underlines the potential of our methodology to address and surmount the complexities inherent in targeting clinically challenging biomolecules.

\section{Conclusion}
%In this paper, we have presented a hybrid quantum-classical algorithm tailored for near-term quantum computers, aimed at discovering novel ligands for specific molecular targets. Our method fuses a Quantum Circuit Born Machine (QCBM) as a prior distribution with a classical Long Short-Term Memory (LSTM) network, alongside a classical molecular filter and reward function. We assessed the efficacy of our hybrid model in two ways. Computationally, we conducted benchmark studies focusing on success rates, where we noted an enhancement in the quality of molecules (reflected as increased success rates) when using the hybrid model as opposed to a solely classical approach. This implies that the quantum-enhanced model possesses the capability to traverse the synthesizable chemical space more effectively than traditional classical generative models, while concurrently yielding molecules with robust binding affinities.\\
We introduce a hybrid quantum-classical algorithm, meticulously crafted for near-term quantum computers, aiming to discover novel ligands for specific molecular targets. Our method ingeniously integrates a Quantum Circuit Born Machine (QCBM) as a prior distribution with a classical Long Short-Term Memory (LSTM) network, enhancing this amalgamation with a reward function designed to foster the generation of drug-like small molecules. We benchmarked our approach through computational evaluations across two distinct sets of tasks. Initially, we engaged in a comparative analysis against classical generative models, focusing on the design of potent small molecule binders for three different proteins, employing the Tartarus benchmarking suite. The results highlight our model's ability to generate high-quality molecules, registering the highest success rate across the tasks, albeit with a marginally lower docking score when juxtaposed with classical approaches.  In addition, we examined the impact of the number of qubits on the modelling of the prior distribution and observed a trend that suggests a roughly linear correlation between the modelling success rate of generating high-quality molecules and the number of qubits involved. These findings suggest a potential advantage of the quantum-enhanced model in navigating the synthesizable chemical space with greater efficacy than its traditional classical counterparts while simultaneously yielding molecules that exhibit strong binding affinities. \\

%Regarding the second validation approach, our model was applied in a practical setting, where we aimed to design inhibitors for the notoriously difficult cancer target, KRAS. For this task, we experimentally synthesized and tested 15 promising ligands generated by both the hybrid and the fully classical models. Our experiments confirmed the inhibitory effect of two molecules on KRAS, validating our models' capability to generate viable new ligands for challenging drug targets. Notably, some of these compounds demonstrated binding affinities in the lower micromolar range, with cell-based assays further confirming their promising \textit{in vitro} effects. The most effective molecules emerged from the hybrid quantum/classical approach, suggesting a superior performance over the fully classical model. Moreover, the distinct molecular candidates produced by the classical and hybrid models highlight their exploration of different regions within the chemical space.\\

Subsequently, we deployed our model in a practical scenario to design inhibitors targeting the notoriously challenging cancer protein, KRAS. In this endeavour, we experimentally synthesized and evaluated 15 promising ligands conceived by hybrid and fully classical models. Our empirical findings substantiated the inhibitory properties of two molecules, ISM061-018-2 and ISM061-22, on KRAS, affirming our models' proficiency in generating viable new ligands for complex drug targets. ISM061-018-2 was identified as a binder to KRAS-G12D, exhibiting a binding affinity of $1.4 \mu M$, and it demonstrated pan-KRAS inhibition. Furthermore, ISM061-018-2 exhibited specificity towards certain mutants, showing pronounced activity, particularly against KRAS G12R and Q61H. Both molecules were derived from our hybrid methodology, indicating their superior efficacy to the fully classical model. In addition, both compounds introduced novel chemotypes distinct from existing KRAS inhibitors, illustrating our hybrid model's capacity to effectively navigate and explore diverse regions within  chemical space.\\

%The results presented do not necessarily indicate that our quantum-enhanced model demonstrated a 'quantum advantage'—that is, delivering outcomes unattainable by classical means within a practical timeframe. Given the modest qubit count (16) utilized, our hybrid algorithm remains amenable to full classical simulation. Furthermore, it is conceivable that alternative classical algorithms could surpass the performance of our quantum approach. Investigating the comparative benefits of our hybrid quantum-classical algorithm against analogous classical counterparts represents a compelling avenue for future research. Points of interest for subsequent inquiry include the scaling behavior in relation to the number of qubits employed, their specific types and interconnections, the impact of quantum noise and errors, and performance benchmarks against leading classical algorithms with respect to success rates. Additionally, alternative metrics beyond success rate, such as the docking scores of ligands, warrant further exploration.\\
While the results showcased are promising, they do not conclusively establish a 'quantum advantage', defined as achieving results beyond the reach of classical methods within a reasonable timeframe. The modest count of 16 qubits in our hybrid algorithm permits simulation on classical platforms, suggesting that state-of-the-art classical algorithms might match or even exceed the efficacy of our quantum-assisted approach. Hence, a critical future research direction involves comprehensively assessing our hybrid quantum-classical algorithm's performance compared to its classical equivalents. Essential factors for this comparative study include analyzing the scalability relative to qubit quantity, the intricacies of qubit types and their interconnections, the influence of quantum noise and errors, and how the algorithm measures up against top-tier classical algorithms in terms of success rates and other crucial metrics like the docking scores of ligands.\\

Our research indicates that current near-term quantum hardware can already be harnessed for practical drug discovery applications, mitigating the need to wait for fully fault-tolerant quantum computers, which may be a decade from fruition. Moreover, since our algorithm uses only 16 qubits within the realm of classical simulation, it shows how quantum computing can catalyze the development of more efficient algorithms for classical hardware. In conclusion, we have introduced a hybrid quantum-classical algorithm surpassing its classical performance counterpart. The modest qubit count used, absent any error correction and with limited connectivity,  hints at the potential of more advanced quantum computers alongside better quantum-classical algorithms for future drug discovery. With an anticipated increase in qubits, improved fidelity, error correction capabilities, and enhanced connectivity, the prospects for quantum computing in drug discovery are a new frontier of computational and experimental science.

\section{Acknowledgements}
We want to thank Alejandro Perdomo-Ortiz, Marta Mauri, Brian Dellabetta, Vladimir Vargas-Calderón, Austin Cheng, Jacob Miller, Mohsen Bagherimehrab, and Kouhei Nakaji for their valuable discussion and support in this research. Additionally, we are thankful to the Defense Advanced Research Projects Agency (DARPA) for their funding (grant number HR0011-23-3-0017), which significantly supports our scientific endeavours.
AK.N. acknowledges funding from the Bio-X Stanford Interdisciplinary Graduate Fellowship (SGIF). A.A.-G. thanks Anders G. Fr{\o}seth for his generous support.  A.A.-G. also acknowledges the generous support of Natural Resources Canada and the Canada 150 Research Chairs program. This research was undertaken thanks in part to funding provided to the Acceleration Consortium of the University of Toronto from the Canada First Research Excellence Fund. 
\newpage
\newpage

\bibliographystyle{unsrt}
\bibliography{references,references_zoteros}
\appendix
\section{Methods}\label{sec:method}

\begin{figure*}
    \centering
    \includegraphics[width=\linewidth]{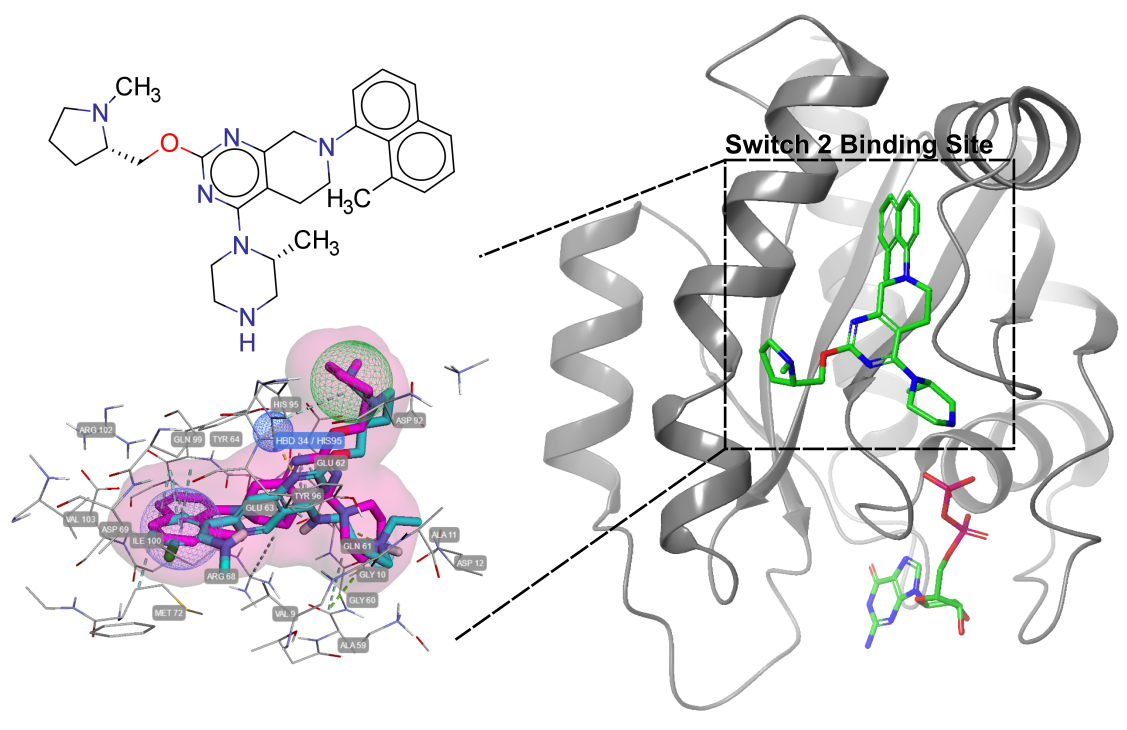}
\caption{\textbf{Depiction of the Pharmacophore Model for KRAS Inhibitors Based on the Co-crystallized Ligand Structure Analyzed with Chemistry42} (PDB: 7EW9 \cite{bank_rcsb_nodate}). Essential pharmacophoric elements are delineated: a blue sphere illustrates a ring system that contributes to structural integrity, a green sphere marks a hydrophobic moiety pivotal for binding affinity, and a cyan sphere signifies a hydrogen bond donor, integral for specificity and interaction with the KRAS protein. The protein structure is displayed on the right, with the pharmacophore interactions within the KRAS Switch 2 Binding Site illustrated on the left.}
    \label{fig:SAR}
\end{figure*}

\begin{table*}[t]
\centering
% \caption{Binding parameters of reference compounds to KRAS G12D.}
\caption{\textbf{Summary of binding kinetics and affinity parameters for reference compounds interacting with the KRAS G12D mutation.} The table presents the on-rate (\(K_{on}\)), off-rate (\(K_{off}\)), and dissociation constants (\(K_d\)) in both kinetic and steady-state measurements, providing a comprehensive overview of the binding dynamics of MRTX1133 and BI-2852 compounds with the target protein.}
\label{tab:binding_parameters}
\begin{tabular}{lcccc}
\hline
% \rowcolor[HTML]{DECBFF} 
% \toprule
\textbf{Compound} & $K_{on} \, (M^{-1}s^{-1})$ & $K_{off} \, (s^{-1})$ & \textbf{Kinetic} $K_d \, (M)$ & \textbf{Steady state} $K_d \, (M)$ \\
% \midrule
\hline
\midrule
MRTX1133 & $1.26 \times 10^{10}$ & $0.0732$ & $5.83 \times 10^{-12}$ & -- \\ 
\midrule
BI-2852   & $6.13 \times 10^{6}$  & $0.0101$ & $1.65 \times 10^{-9}$  & $1.41 \times 10^{-8}$ \\
% \hrule
\end{tabular}
\end{table*}

\begin{table*}[t]
\centering
\caption{\textbf{Dissociation constants (\(K_d\)) for a series of newly generated compounds, illustrating their binding affinities to the KRAS G12D protein.} Five of the twelve compounds evaluated demonstrated discernible binding affinities, with one notable compound achieving binding affinity in the single-digit micromolar range. Contrasting with the reference inhibitors, these compounds exhibit a 'fast off' kinetic profile, which may hold significance for their pharmacodynamic properties. While their affinities are considered weak for immediate therapeutic application, these compounds, synthesized from a small and structurally diverse training set, represent promising scaffolds for further optimization in pursuit of novel KRAS protein inhibitors.}
\label{affinity_genereted_compounds}
\begin{tabular}{lc|cc|cc|ccc}
% \toprule
\hline
% \rowcolor[HTML]{DECBFF} 
\textbf{Compound} & \textbf{K\textsubscript{d} (\(\mu\)M)} &\textbf{Compound} & \textbf{K\textsubscript{d} (\(\mu\)M)} & \textbf{Compound} & \textbf{K\textsubscript{d} (\(\mu\)M)}& \textbf{Compound} & \textbf{K\textsubscript{d} (\(\mu\)M)} \\ \hline \midrule
ISM061-6& 39.1& ISM061-18 & 19.2&ISM061-14& N/A & ISM061-22& N/A\\ \midrule
ISM061-11& N/A& ISM061-18-2& 1.4  &ISM061-15& N/A& ISM061-4   & N/A \\ \midrule
ISM061-13& N/A& ISM061-21  & 56.9 &ISM061-16& N/A& ISM061-24-2& 16  \\ \bottomrule
\hline
\end{tabular}
\begin{tablenotes}
      \small
      \item Note: N/A – No Activity observed.
    \end{tablenotes}
\end{table*}

This section explains the methods and workflow incorporated in our proposed approach, offering a comprehensive understanding of the mechanisms used in our study. Figure~\ref{fig:workflow} illustrates the workflow we employed in our study.

\subsection{Data Acquisition and Pre-processing}
Our preliminary dataset, sourced from Insilico Medicine\footnote{https://insilico.com/}, included approximately 650 data points. These were selectively collated from existing literature, specifically targeting the KRAS G12D mutation (refer to Figure~\ref{fig:workflow}). Given the dataset's limited size, we opted to expand it to improve the robustness of our model during training.  
% Upon analysis, we observed a high degree of similarity among the initial compounds too; this similarity is graphically depicted in Figure~\ref{fig:hitmap_similarity}. Accordingly, we established a workflow to mine the ligands to increase hit expansion.
% \begin{figure}[hbt!]
%   \centering
%   \includegraphics[width=\linewidth]{plots/similarity_matrix_heatmap_viridis.pdf}
%   \caption{\textbf{Chemical Diversity Heatmap of KRAS G12D Ligands: The heatmap depicts the pairwise Tanimoto similarity scores for a collection of ligands targeting the KRAS G12D mutation.} Each square indicates the degree of similarity between pairs of ligands, with the diagonal representing a perfect match (score of 1.0). The dark blue to yellow gradient demonstrates the range from highly similar (yellow) to less similar (dark blue) compounds. Prominent yellow and green-ish clusters suggest groups of ligands with structural similarities, potentially indicative of shared binding modes or pharmacophores critical for interaction with the KRAS G12D variant.}
%   \label{fig:hitmap_similarity}
% \end{figure}

% 

\subsubsection{STONED-SELFIES}\label{appx:stonedselfies}
We utilized the STONED-SELFIES \cite{nigam_beyond_2021} algorithm, available at \href{https://github.com/aspuru-guzik-group/stoned-selfies}{https://github.com/aspuru-guzik-group/stoned-selfies}, to mine our initial set of 650 molecules. For a given molecule in SMILES format, we first randomized the string using RDKit\footnote{https://www.rdkit.org/}. These randomized strings were then converted into SELFIES. Each SELFIES string underwent mutations—in the form of character deletions, replacements, and additions—up to 500 times. Subsequently, the synthesizability and stability of the mutated strings were assessed using Chemistry42. We generated 850k molecules, which served as the training set for our generative models.

\subsubsection{Virtual Screening Process}\label{appx:virtualflow}
VirtualFlow 2.0 \cite{gorgulla_virtualflow_2023} was used to identify additional molecules predicted to bind to KRAS G12D. The adaptive target-guided (ATG) method performed the virtual screening in two stages. In the first stage, the ATG prescreen was performed, in which a spare version of the 69 billion REAL Space from Enamine (version 2022q12) was screened. In the second stage, the most potent tranches of ligands were screened in full, amounting to 100 million ligands. The docking program used was QuickVina 2 \cite{alhossary2015fast}, with exhaustiveness set to 1 in both stages of the screen. The screen was carried out in the AWS (Amazon Web Services) cloud computing platform. The protein structure used in the screen is the PDB structure 5US4 \cite{welsch2017multivalent}, which was prepared before the virtual screen with Schrödinger's Protein Preparation Wizard (addition of hydrogens, protonation state prediction). The docking box was of size 14x14x20 Angstrom. 

\subsection{Quantum Assisted Algorithm}\label{sec:qcbm}
As Figure~\ref{fig:workflow} shows, Our quantum-assisted model is a hybrid algorithm composed of both quantum and classical generative components. The quantum generative model utilizes a Quantum Circuit Born Machine (QCBM) model, while the classical component utilizes a Long Short Term Memory (LSTM) model. Figure~\ref{fig:qegm} illustrates the flowchart of our proposed generative model.

Within this model, we utilized Chemistry42 and a local filter to validate sample generation at each step, which was then employed to train the QCBM model. The QCBM model, a quantum circuit model, was executed on a quantum processing unit. Subsequently, samples from the trained QCBM were fed into the LSTM model, which generated sequences based on these samples. The reward value for each sample was computed at every step using Local filter until epoch 20, and after we selected Chemistry42. This reward value was then used to train our quantum generative model. During the first epoch, no rewards were available, so the algorithm sampled from the untrained QCBM model, designated as $X_{i}$. From the second epoch onward, rewards were computed, allowing us to calculate the \textbf{Softmax} of the rewards for each $X_{i}$, where $i \in [1, N]$. The corresponding pseudocode can be found in Algorithm~\ref{alg:hybrid_v0}.

% \begin{figure}[bt!]
%   \centering
%   \includegraphics[width=\linewidth]{figures/qcbm_lstm_arch_fig2_gen.png}
%   \label{fig:compound_selection}
%     \caption{\textbf{Schematic representation of the hybrid quantum-classical computational model employed for ligand discovery}. The architecture integrates Long Short-Term Memory (LSTM) cells, optimized for processing sequential data, with a Quantum Circuit Born Machine (QCBM). Inputs, represented as $X(i)$, originate from the QCBM and are further processed by the LSTM cells. The iterative training process relies on sample quality assessments facilitated by Chemistry42, which informs the QCBM training. This cyclical validation and feedback mechanism is instrumental in purifying the predictive accuracy of the model.}

% \end{figure}

\begin{algorithm*}[b]
\caption{\textbf{overview of Quantum-Assisted Drug Discovery using an LSTM framework.} This pseudocode details the iterative process, starting with the initialization of the LSTM and QCBM models, followed by the generation and validation of new compounds. Valid compounds are subjected to a reward calculation and probability assessment, which in turn inform the subsequent training of the QCBM (detailed in Algorithm \ref{alg:qcbm_training}). This cycle continues until convergence, illustrating the dynamic interplay between quantum predictions and LSTM-generated compounds underpinned by the Chemistry42 evaluation.}
\label{alg:hybrid_v0}
\begin{algorithmic}[1]
\State \textbf{Initialize:} LSTM, QCBM, filter (Chemistry42)
\State \textbf{Generate} initial samples $X_i$ from QCBM

\While{not converged}
\State \textbf{Train} LSTM with $X_i$
\State LSTM generates a new compound from the current samples $X_i$
\State \textbf{Validate} the new compound with the filter
\If{new compound is valid}
\State \textbf{Compute} rewards for the new compound
\State \textbf{Compute} probabilities $P(X_i)$ for each new compound
\State \textbf{Train} QCBM with $X_i$ and $P(X_i)$
\State \textbf{Generate} new $X_i$ from QCBM
\EndIf
\EndWhile
\end{algorithmic}
\end{algorithm*}

\subsubsection{QCBM Model}
The Quantum Circuit Born Machine (QCBM) model represents a quantum variational generative model, necessitating a classical optimizer to train its parameters. The total count of these parameters is computed with the number of qubits and layers defined in the model (refer to Figure~\ref{fig:qcbm_3layers_v0}). Upon specifying the parameters, denoted as \(\theta_n\), we obtain a quantum state \(\ket{\psi(\theta)}\). Here, each \(\theta_n\) exerts an impact on the wave function, expressed as \(\psi(\theta)\). To optimize these parameters \(\theta\), the model is initially configured with randomly assigned parameters \(\ket{\psi(\theta)}\). These parameters are subsequently calculated throughout the training process. The training of the QCBM model involves minimizing the Exact Negative Log-Likelihood (Exact NLL) loss function.

% To do this, the  Exact NLL function concerning $\theta$ is computed. the optimizer is used to adjust the parameters. This process is repeated until the NLL reaches to the lower value. During the training process, the validity and the reward value for each sample are computed at each step. If the sample is deemed valid, the rewards are used to adjust the probabilities $P(X_i)$ of each new compound. These probabilities are then used to train the QCBM model further.

\subsubsection{Classical Model: LSTM Model}
Long Short-Term Memory (LSTM) networks (see Figure~\ref{fig:workflow}) are employed for a classical part of this architecture. LSTM is simple and has a good record of learning the string pattern in natural language processing for a long time. LSTM networks are specialized Recurrent Neural Networks (RNN) capable of learning long-term dependencies in sequence data \cite{hochreiter_long_1997}. They are particularly useful in applications where the context from earlier parts of the sequence is needed to interpret later parts, such as in natural language processing, time-series forecasting, and more \cite{gers_learning_1999}.
The LSTM architecture consists of a chain of repeating modules called cells. Each cell contains three gates that control the flow of information:

\begin{enumerate}
    \item \textbf{Forget Gate}: This gate decides what information from the cell state should be thrown away or kept. It takes the output of the previous LSTM cell and the current input and passes them through a sigmoid function, outputting a number between 0 and 1 for each number in the cell state. A 0 means ``completely forget this'' and a 1 means ``completely keep this.''
    \item \textbf{Input Gate}: This gate updates the cell state with new information. It has two parts: a sigmoid layer called the ``input gate layer'' and a hyperbolic tangent layer. The sigmoid layer decides what values to update, and the hyperbolic tangent layer creates a vector of new candidate values that could be added to the state.
    \item \textbf{Output Gate}: This gate decides the next hidden state. The hidden state contains information on previous inputs. The hidden state is used to calculate the output of the LSTM and the next hidden state.
\end{enumerate}

The following equations can describe the LSTM's operations:
{\small
\begin{align}
    \text{Forget Gate:} & \quad f_t = \sigma(W_f \cdot [h_{t-1}, x_t] + b_f) \\
    \text{Input Gate:} & \quad i_t = \sigma(W_i \cdot [h_{t-1}, x_t] + b_i) \\
    \text{Candidate Values:} & \quad \tilde{C}_t = \tanh(W_C \cdot [h_{t-1}, x_t] + b_C) \\
    \text{Update Cell State:} & \quad C_t = f_t \cdot C_{t-1} + i_t \cdot \tilde{C}_t \\
    \text{Output Gate:} & \quad o_t = \sigma(W_o \cdot [h_{t-1}, x_t] + b_o) \\
    \text{Update Hidden State:} & \quad h_t = o_t \cdot \tanh(C_t)
\end{align}
}
Where \( \sigma \) is the sigmoid activation function, \( W \) and \( b \) are the weight matrices and bias vectors for each gate, and \( x_t \) is the input at time \( t \) \cite{gers_recurrent_2000}.

Training an LSTM involves optimizing the network's weights and biases to minimize a specific loss function. This is typically accomplished using gradient-based optimization algorithms such as stochastic gradient descent (SGD) or Adam \cite{kingma_adam_2017}. The backpropagation through time (BPTT) algorithm is employed to compute the gradients relative to the loss function, considering the sequential nature of the data \cite{werbos_backpropagation_1990}. The networks are trained using the Adam Optimizer with the Negative Log Likelihood Loss function, and to mitigate overfitting, regularization techniques like dropout are implemented \cite{srivastava_dropout_2014}.
The Negative Log-Likelihood Loss for a single data point is given by:
\begin{equation}
    L(y, \hat{y}) = -\log(\hat{y}_y)
\end{equation}
where \( y \) is the true class label, and \( \hat{y}_y \) is the predicted probability for the true class label \( y \).

The loss for a batch of data is the mean of the individual losses for each data point in the batch:
\begin{equation}
    \mathscr{L} = -\frac{1}{N}\sum_{i=1}^{N}\log(\hat{y}_{y_i})
\end{equation}
Where \( N \) is the number of data points in the batch, \( y_i \) is the true class label for the \( i \)-th data point, and \( \hat{y}_{y_i} \) is the predicted probability for the true class label of the \( i \)-th data point.

In the hyperparameter tuning process, we utilized Optuna\footnote{https://optuna.org/}, an optimization framework, to adjust parameters such as the number of hidden dimensions, embedding dimensions, and layers within the model.
The model presented in this research integrates a deep learning architecture. This architecture is designed to incorporate prior information (samples) into the generative process. Additionally, the model employs Chemistry42 feedback in conjunction with Quantum Circuit Born Machines (QCBM), aiming to enhance its generative accuracy. Figure~\ref{fig:qegm}\textbf{(B)} illustrates the proposed architecture at a cell level. 
% This architecture consists of an LSTM employing temperature-controlled sampling techniques. 
The prior samples are combined with Input samples $ x\prime_t = X(i) \mdoubleplus x_t $ in LSTM cell; This combination has two methods: adding and concatenating samples. LSTM's operations will be updated with 

% \begin{align}
%     \text{Prior Sampling} & \quad x\prime_t = X(i) \mdoubleplus x_t  \quad  OR \quad x\prime_t = X(i) + x_t\\
%     \text{Forget Gate:} & \quad f_t = \sigma(W_f \cdot [h_{t-1}, x\prime_t] + b_f) \\
%     \text{Input Gate:} & \quad i_t = \sigma(W_i \cdot [h_{t-1}, x\prime_t] + b_i) \\
%     \text{Candidate Values:} & \quad \tilde{C}_t = \tanh(W_C \cdot [h_{t-1}, x\prime_t] + b_C) \\
%     \text{Update Cell State:} & \quad C_t = f_t \cdot C_{t-1} + i_t \cdot \tilde{C}_t \\
%     \text{Output Gate:} & \quad o_t = \sigma(W_o \cdot [h_{t-1}, x\prime_t] + b_o) \\
%     \text{Update Hidden State:} & \quad h_t = o_t \cdot \tanh(C_t)
% \end{align}
{\small
\begin{align}
    \text{Prior Sampling} & \quad x'_t = X(i) \mdoubleplus x_t  \quad \nonumber \\
    & \text{OR} \quad x'_t = X(i) + x_t  \\
    \text{Forget Gate:} & \quad f_t = \sigma(W_f \cdot [h_{t-1}, x'_t] + b_f) \\
    \text{Input Gate:} & \quad i_t = \sigma(W_i \cdot [h_{t-1}, x'_t] + b_i) \\
    \text{Candidate Values:} & \quad \tilde{C}_t = \tanh(W_C \cdot [h_{t-1}, x'_t] + b_C) \\
    \text{Update Cell State:} & \quad C_t = f_t \cdot C_{t-1} + i_t \cdot \tilde{C}_t \\
    \text{Output Gate:} & \quad o_t = \sigma(W_o \cdot [h_{t-1}, x'_t] + b_o) \\
    \text{Update Hidden State:} & \quad h_t = o_t \cdot \tanh(C_t)
\end{align}
}
% \begin{figure}[hbt!]
%   \centering
%   \includegraphics[width=0.9\linewidth]{figures/qcbm_lstm_v2.pdf}
%   \caption{\textbf{The integration between LSTM and QCBM samples is showcased.} $X(t)$ denotes a sample from the QCBM model, trained on the quality of samples produced in the preceding step. On the other hand, $x(t)$ signifies the initial data encoded in the SELFIES format. These data sources are merged in one of two methods: either by concatenation or addition. Subsequently, the combined data is inputted into the LSTM cell, represented as $x\prime(t)$.}
%   \label{fig:compound_selection}
% \end{figure}

% During generation, the LSTM network processes prior samples from the QCBM, representing specific constraints. Depending on the requirements, either deterministic behavior through greedy sampling or stochastic behavior through temperature-controlled sampling can be applied. 
To generate samples, the process begins with sampling from the prior, followed by the LSTM network processing these prior samples to generate compounds representations. The compounds will be validated through the Chemistry42 platform, specifically tailored to assess ligand quality for the KRAS G12D mutation. This methodology offers designing ligands targeted at specific proteins.
Moreover, the LSTM model is a classical approach for learning ligand structures and constructing a latent ligand space. The Quantum Circuit Born Machine (QCBM) functions as a prior, guiding the LSTM in the generation of novel ligand samples. This procedure is subjected to an iterative process to enhance the quality of ligands, which is evaluated using the Chemistry42 platform.

\subsubsection{Quantum Generative Model: QCBM Model}
The QCBM is a variational quantum algorithm that utilizes the foundational principles of quantum mechanics, particularly the Born rule, to generate complex and diverse data samples. The core of our QCBM model is a parameterized quantum state $\ket{\psi(\theta)}$, where $\theta$ denotes the parameters, or ansatz, of our quantum circuit. As per the Born rule, given a measurement basis, which is commonly the computational basis in our case, the probability of observing a specific outcome $\ket{x}$ is expressed as $|\langle x|\psi(\theta)\rangle|^2$. 
\\
% The QCBM is trained by employing a Maximum Likelihood Estimation (MLE) variant. We aim to adjust the parameters $\theta$ so that the QCBM-generated distribution closely resembles the target data distribution. This optimization is achieved by minimizing the Exact Negative Log-Likelihood (Exact NLL) loss function, which is formulated as:

% \begin{equation}
% \mathscr{L}(\theta) = -\sum_{x} P_{target}(x) \log(P(x))
% \label{eq:nll}
% \end{equation}

% \begin{equation}
%     \text{NLL}(\theta) = -\sum_{x \in \mathcal{X}} \log p_\theta(x)
%     \label{eq:nll}
% \end{equation}

Training a Quantum Circuit Born Machine (QCBM) involves optimizing the parameters of the quantum circuit to produce a probability distribution that closely approximates the target distribution (Probability that is computed by chemistry Reward values. This process is fundamentally iterative, where the quantum circuit parameters, denoted as \( \theta \), are adjusted in each step to reduce the discrepancy between the generated and target distributions. A classical optimization algorithm recommends adjusting parameters, which operates based on the feedback received from the evaluation of the circuit's output. At each iteration, the quantum circuit is sampled to produce a set of states. These states are then compared against the target distribution, and the difference between them informs the direction and magnitude of parameter adjustments in the quantum circuit. This iterative process continues until the distribution generated by the QCBM closely aligns with the target distribution or until a predefined convergence criterion is met.

In the context of QCBM training, the Exact Negative Log-Likelihood (Exact NLL) functions as the primary loss function, providing a quantitative measure of the difference between the distributions. The Exact NLL for a QCBM is the negative sum of the logarithms of the probabilities the quantum circuit assigns to the states in the training dataset. Mathematically, this is represented as \( \text{NLL}(\theta) = -\sum_{x \in D} \log p_\theta(x) \), where \( D \) is the set of data points, and \( p_\theta(x) \) is the probability of observing state \( x \) under the current parameters \( \theta \) of the quantum circuit. Minimizing the NLL involves adjusting \( \theta \) such that the quantum circuit's output distribution increasingly resembles the empirical distribution of the data. This optimization is typically carried out using gradient-based methods or other heuristic techniques suited to the quantum computing context. In our project, we used \text{COBYLA} for our optimizer. As the NLL decreases, the fidelity of the QCBM in modeling the target distribution correspondingly increases, indicating successful training of the quantum model.

\begin{figure*}[t!]
  \centering
  \includegraphics[width=0.9\linewidth]{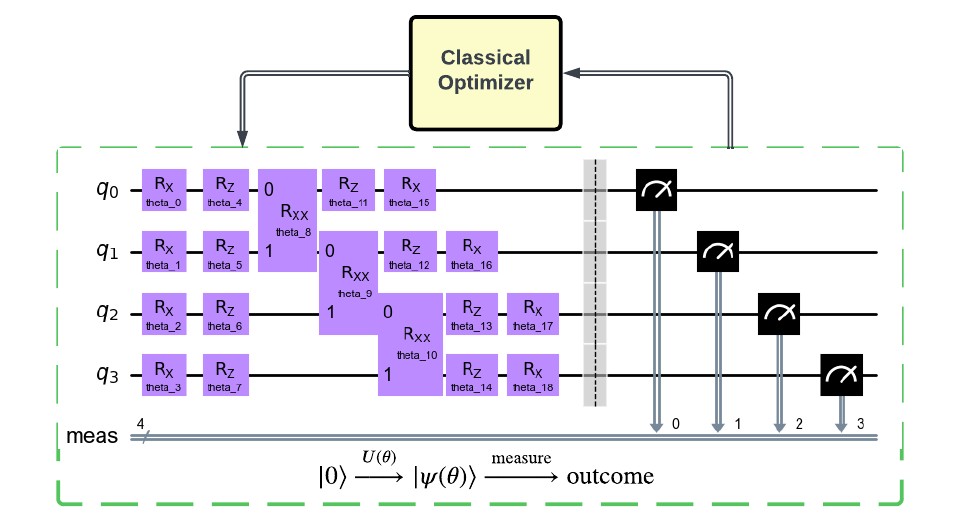}
  % \caption{Illustration of the Quantum Generative model used in our research. It is a three-layer Quantum Circuit Born Machine (QCBM) implemented with four Qubits.}
\caption{\textbf{Schematic representation of the Quantum Circuit Born Machine (QCBM)} implemented in our numerical experiments, illustrating a variational quantum circuit with a configuration of three layers and four qubits. In practice, our numerical experiments utilized a system with 16 qubits. The depicted quantum gates, including parameterized rotations (Rx, Rz) and entangling CNOT gates, are orchestrated to evolve the initial state $|0\rangle$ into a complex quantum state $|\psi(\theta)\rangle$. The outcome is measured, and the resulting data are used by the classical optimizer to iteratively refine the parameters $\theta$, thus leading the circuit towards an optimal solution for ligand generation.}
% , enabling the discovery of novel ligands with desired properties.}
  \label{fig:qcbm_3layers_v0}
\end{figure*}

Figure~\ref{fig:qcbm_3layers_v0} shows the QCBM architecture and illustrates its associated ansatz. We used linear topology for our project. Our QCBM model is built with 16 qubits and 4 layers, and we had 96 parameters to optimize in total. The initial probability of the samples, $P(X(i))$, is computed based on the rewards returned by the Chemistry42 model. These reward-based probabilities are then passed through a \texttt{Softmax} function to ensure they are normalized and fall within the range of 0 to 1. The resulting values serve as the "true" probabilities of the samples and are used as the target values during the model's training process.

\begin{algorithm*}[b!]
% \caption{The pseudocode for the training process of the QCBM model is as follows}
\caption{\textbf{pseudocode outlining the training regimen for the Quantum Circuit Born Machine (QCBM) model.} This process delineates the iterative optimization of the QCBM parameters.}

\begin{algorithmic}[1]
\State \textbf{Initialize:} QCBM model with a certain number of qubits and layers
\State \textbf{Set:} Parameterized quantum state $\ket{\psi(\theta)}$

\While{not converged}
\State \textbf{Compute} exact negative log-likelihood (exact NLL) loss function
\State \textbf{Compute} gradient of exact NLL with respect to $\theta$
\State \textbf{Adjust} parameters using an optimizer
\State \textbf{Validate} the sample and compute its reward value
\If{sample is valid}
\State \textbf{Compute} rewards for the sample
\State \textbf{Adjust} probabilities $P(X_i)$ based on the rewards
\State \textbf{Train} QCBM model with adjusted probabilities
\EndIf
\EndWhile
\end{algorithmic}
\label{alg:qcbm_training}
\end{algorithm*}

\subsection{Benchmark Setup}\label{appx:benchmark}

Our benchmark uses both classical and quantum hardware. Our classical computational setup is based on a cluster equipped with GPU nodes. This cluster consists of two GPU nodes, each with specific features. These features include two AMD EPYC 7V13 64-core Processors, resulting in a total of 128 CPU cores per node. In addition, each node is equipped with 512 GB of RAM. The nodes also contain eight AMD Instinct™ MI100 GPUs, each with a GPU RAM of 32GB. For the classical training, we utilized four of these GPUs in parallel (i.e., one GPU node). Furthermore, we used a Nvidia GPU (RTX3090Ti) to facilitate our classical quantum simulations. For the quantum hardware setup, we employed the Guadalupe quantum system, equipped with 16 Qubits and a Falcon r4P processor type.
% \ref{fig:ibmq}
Our Quantum Circuit Born Machine (QCBM) model, accompanied by an error correction circuit, was executed on this quantum processor.

Regarding software, we utilized several packages provided by Zapata AI under the Qml core agreements \footnote{https://docs.orquestra.io}. We implemented our Variational Quantum Circuit and classical LSTM model using the Qml Core Python package. We used the STONED-SELFIES and VirtualFlow 2.0 packages to prepare a diverse dataset. Additionally, we employed RDkit and Insilico APIs to compute the reward value and conduct some post-processing analyses.
The QCBM model underwent a training regimen spanning 30 epochs. In contrast, the LSTM model was trained over a total of 40 epochs.

We utilized the Optuna platform to optimize the hyperparameters in the benchmarking. We ran Optuna tuning for 100 trials for each model, to determine the optimal number of QCBM layers, LSTM layers, and embedding dimensions. Additionally, we tuned the sampling temperature, which defines the balance between determinism and stochasticity in the model, particularly between the prior input and the LSTM output.

\subsection{SPR Conditions}\label{appx:SPR}
A Biacore 8K system was used for all experiments. For preliminary compound screening, N-terminal biotinylated KRASG12D protein (synthesized by VIVA Biotech (Shanghai) Ltd, purity $\geq$ 95\%) was captured on a Sensor Chip SA (GE Healthcare) at a density of about 2000RU. Protein immobilization was done using 1x HBS-EP+, 2mM TCEP, 2\% DMSO as a running buffer. Protein was injected for 70s at a flow rate of \(5\,\mu L/\text{min}\). The protein concentration was \(5\,\mu\text{g/mL}\). We performed an initial screening of compounds prepared samples by serial 2-fold dilutions from \(200\,\mu M\) to \(0.39\,\mu M\) in 1x HBS-EP+, 2mM TCEP, 2\% DMSO. Samples were injected for 60s at a flow rate of \(30\,\mu L/min\) and dissociation time was 180s. A Biacore 8K machine was used to carry out the SPR experiments and subsequent data analysis.

\subsection{MaMTH-DS Dose-Response Assays}
MaMTH-DS FLP HEK293 reporter cell lines\cite{saraon_drug_2020} stably expressing KRAS (WT or mutant), HRAS, NRAS or EGFR triple mutant L858R/T790M/C797S bait alongside RAF1 (for RAS baits) or SHC1 (for EGFR) preys were seeded into 384-well white-walled, flat-bottomed, tissue-culture treated microplates (Greiner \#781098) at a concentration of 100,000 $/\mathrm{mL}$ ( $50 \mu \mathrm{L}$ total volume/well) in DMEM/10\% FBS/1\% Pen-Strep media. Seeding was performed using a MultiFlo-FX multi-mode liquid dispenser (BioTek). Plates were left at room temperature for 30-60 minutes following seeding before transfer to a Heracell 150i incubator (Thermo) and growth at $37^{\circ} \mathrm{C} / 5 \%$ CO2 for 3 hours. After growth $10 \mu \mathrm{L}$ of DMEM/10\%FBS/1\% Pen-Strep supplemented with $3 \mu \mathrm{g} / \mathrm{mL}$ Tetracycline (to induce bait/prey expression; BioShop, TET701) and $60 \mathrm{ng} / \mathrm{mL}$ EGF (to stimulate RAS signaling; Sigma \#E9644) was added to each well via multichannel pipette. 
As appropriate, $6 \mathrm{X}$ concentration of drug (or DMSO only) was also included in the media, with all lower concentrations produced via serial dilution starting from the highest concentration solution. Plates were then grown overnight ( $18-20$ hours) at $37^{\circ} \mathrm{C} / 5 \% \mathrm{CO}_2$. Luciferase assay was performed the next day using $10 \mu \mathrm{L}$ of $20 \mu \mathrm{M}$ native coelenterazine substrate (Nanolight \#303) per well. Luminescence was measured using a Clariostar plate reader (BMG Labtech) with a Gain of 3200-3800 and a 1 second integration time. All data analysis was performed using Microsoft Excel and GraphPad Prism. Curve fits were performed in Prism (non-linear regression, log(inhibitor) vs. response, variable slope (4 parameters) with bottom constrained to zero, except for EGFR-SHC1 in the presence of ISM061-22, for which no constraint was applied).

\subsection{Cell Viability Assay}
MaMTH-DS FLP HEK293 reporter cell lines\cite{saraon_drug_2020} stably expressing KRAS (WT or G12V mutant) bait alongside RAF1 prey were seeded into 96-well white-walled, \(\mu\)CLEAR\textsuperscript{\textregistered}
% µCLEAR® 
flat-bottomed, tissue-culture treated plates (Greiner \#655098) at 40 000 cells/well in DMEM/10\% FBS/1\% Pen-Strep media ($60 \mu\mathrm{L}$ total volume/well). Seeding was performed using a MultiFlo-FX multi-mode liquid dispenser (BioTek).
Plates were left at room temperature for 30-60 minutes following seeding before transfer to a Heracell 150i incubator (Thermo) and growth at $37^{\circ} \mathrm{C}$ / 5\% CO2 for 3 hours. After growth $30 \mu\mathrm{L}$ of 3X concentration of the drug (or DMSO only) in DMEM/10\%FBS/1\% Pen-Strep media were added to wells, with all lower concentrations produced via serial dilution starting from the highest concentration solution (final drug concentration $30 \mu\mathrm{M}$ to $123 \mathrm{nM}$). 
$37^{\circ} \mathrm{C} / 5 \%$
Plates were then grown overnight (18-20 hours) at $37^{\circ} \mathrm{C}$/ 5\% CO2. Effect of the drug on cell viability was assessed by CellTiter-Glo\textsuperscript{\textregistered}
% CellTiter-Glo® 
Luminescent Cell Viability Assay from Promega (\#G7570). $90 \mu\mathrm{L}$ of the CellTiter-Glo\textsuperscript{\textregistered} reagent were added directly into each well following 30 min equilibration of the plate at room temperature. Contents of the wells were mixed on an orbital shaker for 2 min, and plates were then incubated at room temperature for 10 minutes to stabilize luminescent signal. Luminescence was measured using a Clariostar plate reader (BMG Labtech) with a Gain of 3600 and a 1 second integration time. Values represent the mean $\pm$S.D. of three replicates for each tested drug concentration. All data analysis was performed using Microsoft Excel and GraphPad Prism.\\

% Overall, these live-cell experimental results further highlight the efficacy of our methodology for rapidly detecting small molecule candidates with biological activity against challenging targets of clinical importance.
% \newpage
% \includepdf[pages=-]{supplement_v3.pdf}
% \includegraphics[page=1]{supplement_v3.pdf}

% \includegraphics[page=1, width=\textwidth]{supplement_v3.pdf}
% \newpage
% \includegraphics[page=2, width=\textwidth]{supplement_v3.pdf}
% \newpage
% \end{landscape}
% \includepdf[pages=-, landscape=false]{supplement_v3.pdf}
% \end{sideways}
% \end{rotate}
\end{document}